\documentclass[journal,twoside,web]{ieeecolor}
\usepackage{generic}
\usepackage{cite}
\usepackage{amsmath,amssymb,amsfonts}
\usepackage{algorithmic}
\usepackage{graphicx}
\usepackage{textcomp}
\usepackage{svg}
\usepackage{hyperref}
\usepackage{booktabs, tabularx}
\usepackage[referable]{threeparttablex}

\usepackage{xcolor}
\definecolor{newcolor}{rgb}{.8,.349,.1}
\definecolor{brightcerulean}{rgb}{0.11, 0.62, 0.74}
\newcommand*{\revhl}{\textcolor{black}}

\usepackage{caption}
\usepackage[math]{cellspace}
\usepackage{array}
\usepackage{eqparbox}
\usepackage{url}
\usepackage{hyperref}
\hypersetup{
    colorlinks=true,
    linkcolor=blue,
    filecolor=magenta,      
    urlcolor=cyan,
    pdftitle={HST},
    }
    
\begin{document} 
    \title{COVID-19 Detection from Respiratory Sounds \\with Hierarchical Spectrogram Transformers}
  \author{Idil~Aytekin,
      Onat~Dalmaz,
      Kaan~Gonc,
      Haydar Ankishan,
      Emine U Saritas,
      Ulas Bagci,
      Haydar Celik,
      \\ and Tolga \c{C}ukur$^*$, \IEEEmembership{Senior Member}\vspace{-1.25cm}
  \thanks{This work is partially supported by NIH grants R01-CA246704 and R01-CA240639, and by TUBA GEBIP 2015, BAGEP 2017, and TUBITAK BIDEB 2022 awards. I. Aytekin and O. Dalmaz contributed equally to this work. (Corresponding author: Tolga Çukur).}
  \thanks{I. Aytekin, O. Dalmaz, E.U. Saritas, and T. Çukur are with the Dept. of Electrical and Electronics Engineering, and the National Magnetic Resonance Research Center, Bilkent University, Ankara, Turkey (e-mail: \{idil@ug, onat@ee, saritas@ee, cukur@ee\}.bilkent.edu.tr). K. Gonc is with the Dept. of Computer Engineering, Bilkent University, Ankara, Turkey (kaan.gonc@bilkent.edu.tr). H. Ankishan is with the Stem Cell Institute, Ankara University, Ankara (ankishan@ankara.edu.tr). U. Bagci is with the Dept. of Radiology, ECE and BME, Northwestern University (ulas.bagci@northwestern.edu). H. Celik is with the Children's National Hospital, Washington, DC (haydari@gmail.com).}}

\maketitle

\begin{abstract}
Monitoring of prevalent airborne diseases such as COVID-19 characteristically involves respiratory assessments. While auscultation is a mainstream method \revhl{for preliminary screening of disease symptoms}, its utility is hampered by the need for dedicated hospital visits. Remote monitoring based on recordings of respiratory sounds on portable devices is a promising alternative, which can assist \revhl{in early assessment of COVID-19 that primarily affects the lower respiratory tract}. In this study, we introduce a novel deep learning approach to distinguish patients with COVID-19 from healthy controls given audio recordings of cough or breathing sounds. The proposed approach leverages a novel hierarchical spectrogram transformer (HST) on spectrogram representations of respiratory sounds. HST embodies self-attention mechanisms over local windows in spectrograms, and window size is progressively grown over model stages to capture local to global context. HST is compared against state-of-the-art conventional and deep-learning baselines. Demonstrations on crowd-sourced multi-national datasets indicate that HST outperforms competing methods, achieving \revhl{over 83\%} area under the receiver operating characteristic curve (AUC) in detecting COVID-19 cases.
\end{abstract}


\bstctlcite{IEEEexample:BSTcontrol}

\section{Introduction}
\label{sec:intro}  
Auscultation is a primary step in preliminary assessment of subjects for symptoms of respiratory disorders \cite{falk2010spectro}. Assessment via stethoscope is non-invasive and inexpensive, but it must be performed by a healthcare professional during a hospital visit. As such, early-stage or continuous monitoring of symptoms might not be feasible across broad populations \cite{sun2020clinical}, as experienced during the COVID-19 pandemic that has significantly disrupted access to healthcare facilities across the globe \cite{zhang2020covid}. A promising solution is remote monitoring of respiratory sounds based on audio recordings captured via portable equipment such as mobile or wearable devices \cite{aliverti2017wearable}. While COVID-19 is clinically diagnosed with reverse-transcription polymer chain reaction (RT-PCR) tests \cite{bai2020presumed} and/or radiological imaging \cite{harmon2020artificial,9254002,9178424}, economic and time costs of these lab-administered procedures restrict patient access and elicit backlogs during periods of high transmission \cite{oh2020deep,9194240,zhu2021gacdn}. Remote screening of respiratory sounds can assist in preliminary assessment and risk stratification for potential COVID-19 cases under low-resource settings. By more-informed resource allocation, it can facilitate early referrals and timely interventions to deteriorating patients to help contain the spread of disease \cite{goel2021automatic}.

Auditory screening of respiratory disorders relies on the prevalence of disease-specific features in respiratory sounds \cite{rocha2019open}. Diverse pathology can be encountered in respiratory disorders ranging from inflammation and obstruction to consolidation and pleural effusion. While some commonalities exist among diseases, precise characteristics of pathology including location and severity typically show disease-specific patterns \cite{jackson2020use,dai2020ct}. Imaging studies report that common lung pathology in non-COVID-related pneumonia has central-peripheral distribution, air bronchograms, and pleural enlargement/effusion; whereas COVID-related pneumonia frequently elicits lower-peripheral distribution, enhanced ground glass opacity and vascular enlargement \cite{bai2020performance, zhao2020importance}. These pathomorphological changes have been associated with increased prevalence of adventitious respiratory sounds in COVID-19 such as coarse breathing, wheezes, and crackles \cite{wheezes_crackles}, which may carry distinctive cues compared to respiratory sounds in other diseases such as asthma, chronic obstructive pulmonary disease (COPD), bronchitis and pertussis \cite{rudraraju2020cough,dai2020ct}.

\begin{figure*}[t]
   \begin{center}
   \begin{tabular}{c}
   \includegraphics[width=0.65\textwidth]{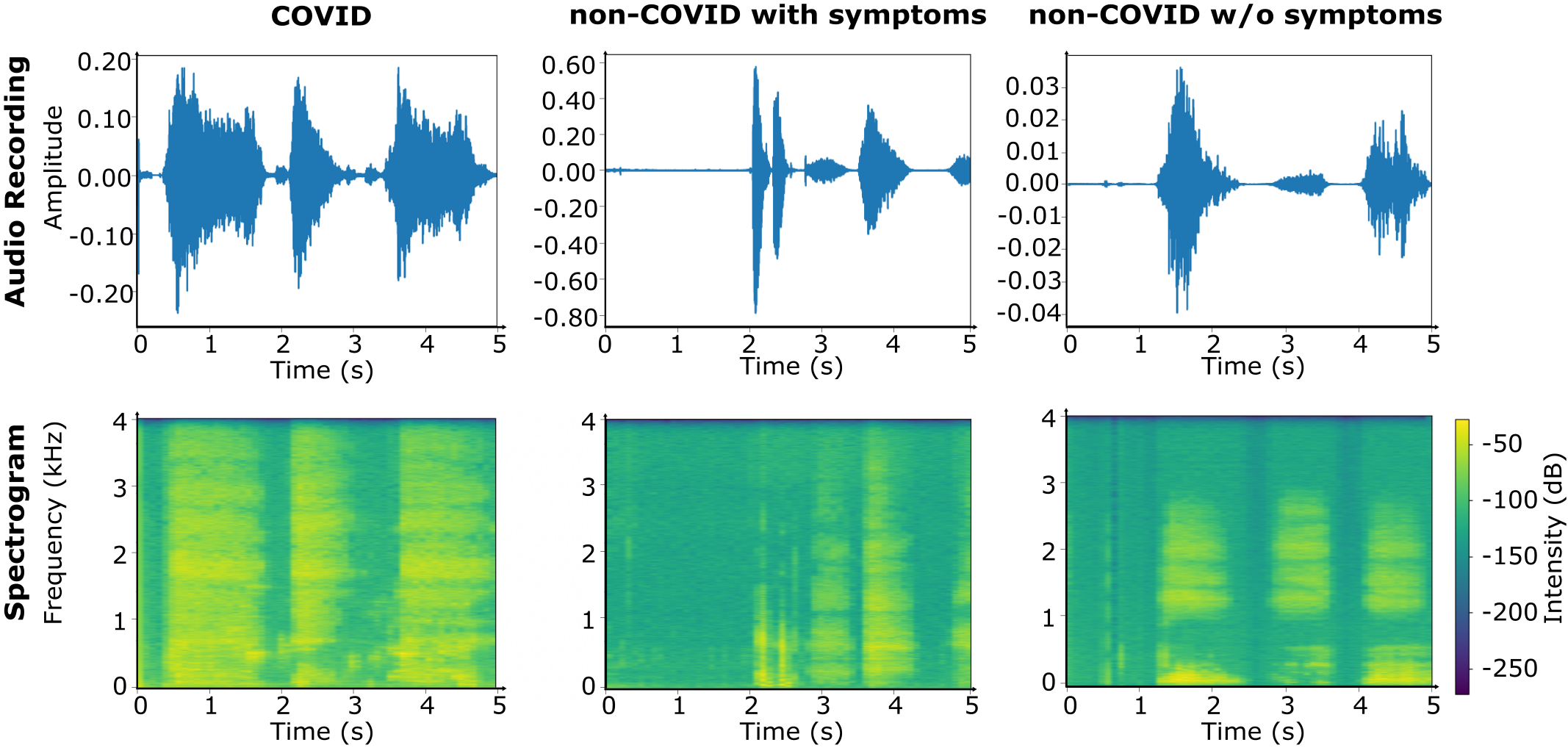}
   \end{tabular}
   \end{center}
   \caption
   { \label{fig:samples} 
Audio recordings of breathing sounds depicted in time-domain (top row) and corresponding spectrogram representations (bottom row). Representative samples are displayed for a COVID case, a non-COVID case with symptoms, and a non-COVID case without symptoms.}
\end{figure*} 

Literature suggests that normal respiratory sounds resemble filtered noise with a typical frequency range of 100-1500 Hz during inspiration/expiration segments of the respiration cycle, and are nearly inaudible during inter-segment intervals \cite{bohadana2014fundamentals}. Meanwhile, coarse breathing and coarse crackles common in COVID-19 resemble dampened sinusoids over a broad frequency range (100-5000 Hz) and extended duration (1-2 s), and wheezes resemble repeated sinusoids of low frequency ($\sim$100 Hz) and short duration ($\sim$100 ms) \cite{bohadana2014fundamentals,wheezes_crackles}. Representative samples of breathing sounds for COVID and non-COVID cases are displayed in Fig. \ref{fig:samples}. The non-COVID case without symptoms follows a cyclic pattern with moderate frequency range and relatively silent inter-segment intervals. The non-COVID case with symptoms shows a degree of irregularity in time and frequency distribution, and modest elevation of intensity at higher frequencies. In comparison, the COVID case shows a uniformly spread intensity distribution across both time and frequency dimensions, relatively stronger intensity at higher frequencies during inspiration/expiration (suggesting coarse breathing/crackles), and brief increases in intensity at low frequencies during inter-segment intervals (suggesting wheezes). Corroborating findings in literature, such apparent and other fine-grained differences in time-frequency characteristics bring forth the possibility of automated screening to identify potential COVID-19 cases.

Given recordings of respiratory sounds, remote screening leverages an algorithm to automatically infer respiratory conditions. Previous studies have successfully applied machine learning (ML) algorithms to detect a broad spectrum of conditions including bronchitis, bronchiolitis, emphysema, pertussis, pneumonia, rale and rhoscus \cite{pramono2016cough,aykanat2017classification,perna2019deep,bales2020can,basu2020respiratory}. Commonly, summary descriptors of audio data were extracted such as Mel-frequency cepstral coefficients (MFCC), tonal and chroma features \cite{srivastava2021deep}. Classifiers were then built via traditional methods such as support vector machines (SVM) and logistic regression, or via network models such as convolutional neural networks (CNN) \cite{aykanat2017classification} and recurrent neural networks (RNN) \cite{perna2019deep,pham2020robust}. Recent studies have followed similar approaches for COVID-19 detection from cough, breathing or speech sounds \cite{despotovic2021detection,mohammed2021ensemble,pal2021pay,chowdhury2022machine}. Various sound descriptors were extracted including MFCC, tonal, chroma, spectral contrast, glottal flow and spectrogram features \cite{deshmukh2021interpreting,mohammed2021ensemble,dash2021detection,chowdhury2022machine}. Either traditional classifiers such as SVM, decision trees and random forests \cite{imran2020ai4covid,brown2020exploring,despotovic2021detection}, or network models including CNNs or RNNs \cite{perna2019deep,pham2020robust,bagad2020cough,laguarta2020covid} were then used for detection.

Despite the potential of learning-based approaches for preliminary COVID-19 screening, there remain avenues for further technical improvement. Many prior studies have employed shallow models ––based on traditional classifiers or neural networks–- in conjunction with hand-crafted audio features. While this approach mitigates model complexity, latent representations in shallow models might have limited power to capture the rich information in respiratory sounds \cite{brown2020exploring}. Few studies have considered additional input features extracted from pre-trained networks, which can be suboptimal compared against task-specific features derived via end-to-end learning \cite{despotovic2021detection}. Other studies have reported elevated detection performance with deep CNN or RNN models that capture a hierarchy of latent representations. Yet, CNNs perform local filtering with compact kernels so they show limited sensitivity to long-range contextual features \cite{vaswani2017attention}. Although RNNs can improve capture of temporal context, serial processing of input sequences can introduce computational burden and compromise feature learning over long time scales \cite{vaswani2017attention}.

In this study, we introduce a novel deep learning method to automatically screen COVID-19 symptoms with short recordings of respiratory sounds. The proposed method first converts respiratory sounds onto a time-frequency (i.e., spectrogram) representation, and then classifies disease from spectrogram features using a novel hierarchical spectrogram transformer (HST). Unlike prior models that use a compact set of knowledge-based features, HST leverages a comprehensive characterization of respiratory sounds through high-resolution log-spectrogram features. Compared to CNNs, HST leverages self-attention mechanism for improved sensitivity to long-range context in spectrograms. Compared to RNNs and vanilla transformers, HST leverages a patch-based approach where the time-frequency extent of attended regions is progressively increased for computational efficiency. 

\vspace{-0.25cm}
\subsection*{{Main Contributions:}}
\begin{itemize}
  \item We introduce \revhl{a novel hierarchical spectrogram transformer for screening COVID-19 symptoms} from audio recordings of respiratory sounds.
  \item \revhl{The proposed transformer model} progressively captures local to global context in spectrogram representations for enhanced efficiency. 
  \item We demonstrate improved performance in COVID-19 detection against state-of-the-art baselines including traditional, CNN, RNN, transformer, and ensemble methods.    
\end{itemize}

\section{Related Work}

\subsection{Shallow Classifiers}
A first group of studies have proposed to detect COVID-19 via shallow models based on either traditional ML methods (e.g., SVM, random forests, decision trees), or neural networks of limited depth. In \cite{imran2020ai4covid}, MFCC features of cough sounds were analyzed with SVM to separate COVID-19 from other respiratory infections. 
In \cite{mohammed2021ensemble}, a diverse set of audio features including the power spectrum, Mel spectrum, chroma, tonal, and MFCC features were extracted from cough sounds, and analyzed with SVM and logistic regression models.
In \cite{ponomarchuk2022project}, cough, breath, and voice recordings were analyzed with an ensemble of shallow CNN, gradient boosted trees and logistic regression models given Mel-spectrogram \revhl{and cochleagram features along with network-based features from a pre-trained convolutional architecture for audio data (VGGish) \cite{vggish}}.
In \cite{pal2021pay}, hand-crafted features of cough sounds including MFCC, log energy and entropy were analyzed via a shallow five-layer model with gated linear units. Cough features were augmented with symptomatic and demographic features for improved performance. 
In \cite{chowdhury2022machine}, MFCC, tonal, chroma, spectral contrast, spectrogram features of cough sounds were input to an ensemble of decision-tree, logistic regression, random forest, boosting and multi-layer perceptron (MLP) models. 
In \cite{brown2020exploring} and \cite{erdogan2021}, COVID-19 screening was proposed based on spectrogram and wavelet features of respiratory sounds concatenated with network-based features obtained from pre-trained VGGish or CNN models. The compiled features were then processed with an SVM or logistic regression model.  
In \cite{dash2021detection}, a new bio-inspired cepstral feature was proposed for COVID-19 detection, and demonstrated on speech, breathing, or cough sounds with SVM. 
In \cite{despotovic2021detection} acoustic and scattering features were augmented with pre-trained VGGish features of cough, breath, speech sounds. Decision trees, random forests or MLP models were then used. 
A common aspect of these previous studies is that they leveraged shallow models of low complexity to process either hand-crafted features or task-agnostic features from pre-trained networks. While this approach can improve learning behavior on limited datasets, the resultant classifiers are deprived from a diverse hierarchy of task-specific latent features that could be captured via end-to-end deep learning. 

\subsection{Deep Classifiers}
A second group of studies have instead considered deep models typically based on CNN or RNN architectures for COVID-19 detection. 
In \cite{imran2020ai4covid}, a CNN was used to predict COVID-19 given MFCC features of cough sounds.
In \cite{bagad2020cough}, short-time magnitude spectrogram of cough sounds were classified via a CNN.
In \cite{mohammed2021ensemble}, power spectrum, Mel spectrum, chroma, tonal, and MFCC features of cough sounds were classified using an ensemble of pre-trained CNN and traditional ML models. 
In \cite{laguarta2020covid}, an ensemble of CNNs was used given as input MFCC features of cough sounds. Three different pre-trained CNNs were used to extract features related to lung and respiratory tract, vocal cord, and sentiment information. 
In \cite{chowdhury2021qucoughscope} and \cite{schuller2020detecting}, spectrogram features of cough and breathing sounds were analyzed with an ensemble of CNNs. 
In \cite{coppock2021end}, spectrogram features of cough and breathing sounds were analyzed via a CNN. 
In \cite{deshmukh2021interpreting}, glottal flow features of speech sounds were extracted and classified via a CNN. \revhl{In \cite{deepshufnet}, augmented mel-spectrogram features of breathing sounds were analyzed via a CNN.} Although CNN models are powerful in capturing local features in time-frequency representations of audio data, the inductive bias introduced by filtering with local kernels limits sensitivity for long-range contextual features.

\begin{figure*}[t]
   \begin{center}
   \begin{tabular}{c} 
   \includegraphics[width=0.65\textwidth]{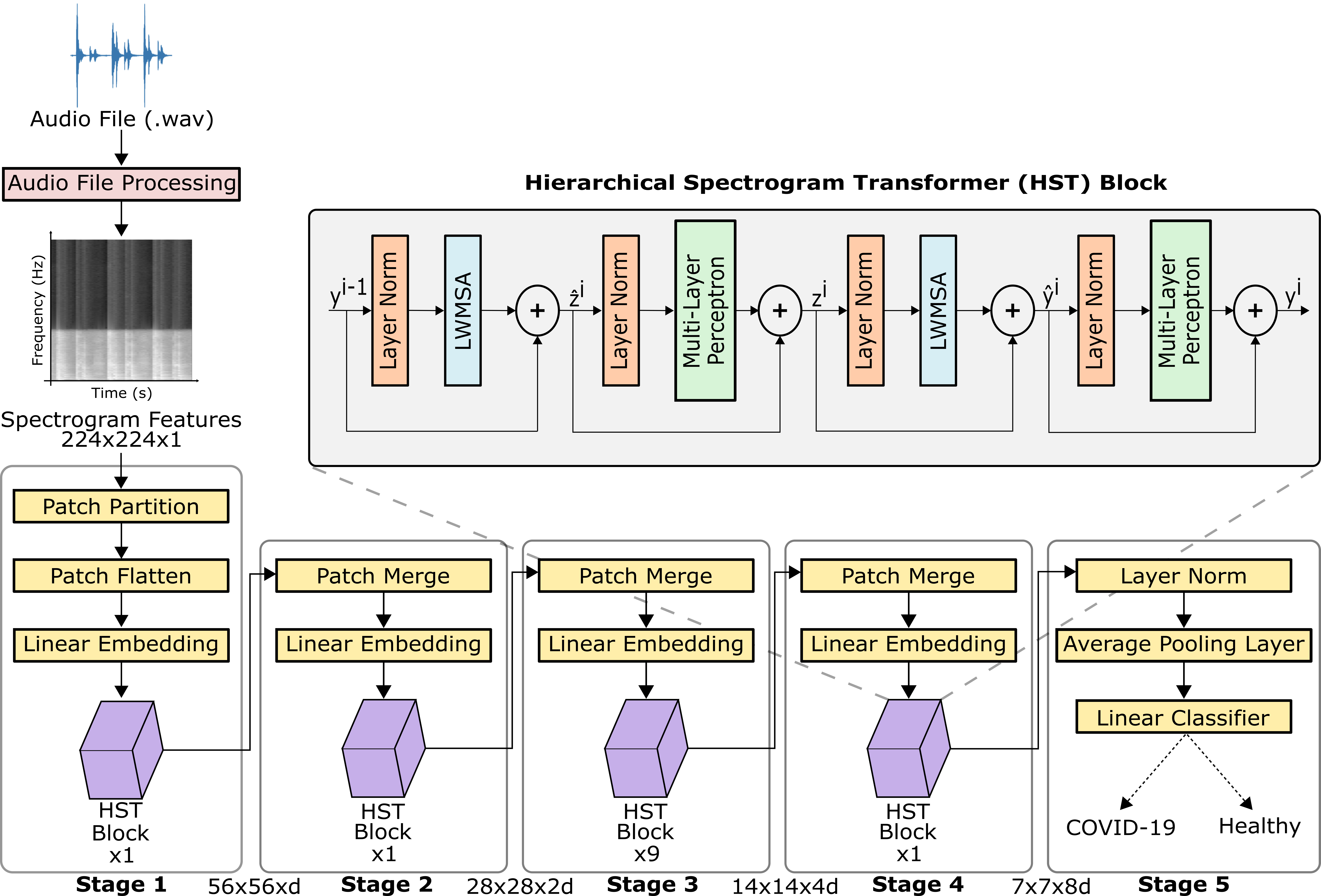}
   \end{tabular}
   \end{center}
   \caption
   { \label{fig:architecture} 
Overall architecture for the hierarchical spectrogram transformer (HST) model for COVID-19 detection based on audio recordings of bodily sounds. To derive the model input, audio recordings are first mapped onto a spectrogram that captures a two-dimensional time-frequency representation. The spectrogram is then processed with HST that employs a hierarchical patch-based attention mechanism to capture local to global context. HST is a five-stage model where the first four stages derive latent audio representations with a cascade of local-windowed transformer blocks. The spectrotemporal resolution of the feature maps is progressively lowered across these stages, while the embedding dimensionality is increased. The last stage maps latent representations onto a classification output via cross-entropy loss.}
\end{figure*} 

Several recent studies have instead proposed architectures designed for sequence modeling to improve capture of temporal context. In \cite{pahar2021covid}, MFCC, chroma and spectral features of cough sounds were analyzed with a long-short-time-memory (LSTM) model. 
In \cite{hassan2020covid}, spectral centroid, spectral roll-off, zero-crossing rate, MFCC, first and second derivatives of MFCC features of cough, breathing, and voice sounds were analyzed with an LSTM.
In \cite{karas21_interspeech}, both hand-crafted and learning-based features from spectrograms of cough sounds were analyzed with an ensemble of SVMs and LSTMs.
In \cite{khriji2021covid}, an LSTM was proposed to detect COVID-19 from MFCC, power spectrum, and filter bank features of cough, breath and sneeze sounds.
In \cite{kamble2022exploring}, cepstral coefficients and filter bank features of breathing, cough, and speech sounds were analyzed with a bidirectional LSTM.
In \cite{hamdi2022cnnlstm} and \cite{vjietdeep}, hybrid CNN-LSTM architectures were proposed for COVID-19 detection, which received as input MFCC or spectrogram features of cough sounds. 
In \cite{pinkas2020sars}, network-based features of speech and cough sounds were inspected via a gated recurrent unit (GRU) model that also received clinical features for improved accuracy.
While RNNs have demonstrated success in capturing signal correlations over long time scales, they can suffer from suboptimal learning due to vanishing gradients. 

Here, we introduce \revhl{a novel hierarchical transformer model} for COVID-19 detection from spectrogram features of respiratory sounds. \revhl{Few recent studies have independently considered transformer models for analyzing respiratory sounds in COVID-19 patients. In \cite{pinkas2020sars}, sound embeddings extracted from a pre-trained transformer are analyzed with an RNN for classification.} Note that \cite{pinkas2020sars} use a vanilla transformer of quadratic complexity, and perform task-agnostic pre-training to extract sound embeddings that are frozen during the classification stage. Instead, HST employs a hierarchical patch-based attention to progressively capture local to global context while alleviating computational burden, and it is trained end-to-end for disease classification resulting in task-specific representations. \revhl{In \cite{hmt}, spectrogram, MFCC and other clinical features are analyzed simultaneously with a nested multi-modal transformer. The nested transformer splits audio spectrograms into non-overlapping patches of growing size across a hierarchy. Yet, the nested transformer independently processes single patches in each stage, only permitting interactions among neighboring patches via pooling operations during block aggregation. In contrast, HST uses cyclic shifts to split spectrograms into partly overlapping patches. It further leverages self-attention operations to enable interactions between a broader set of patches at each stage, which can improve sensitivity to global context and model performance.}

\section{Methods}

\subsection{Experimental Datasets}
\label{sec:dataset}
\revhl{We performed demonstrations on two public datasets containing audio recordings of respiratory sounds: Cambridge \cite{brown2020exploring} (breathing and cough), and COUGHVID (cough) \cite{orlandic2021coughvid}. \revhl{In these datasets, vocal respiratory sounds were recorded via microphones on cellular phones or personal computers. Participants were instructed to record in a silent environment, and breathe deeply through their mouth or cough directly to the microphones held within arm's length}.}

\subsubsection{Cambridge Dataset}
The dataset included three subject groups \cite{brown2020exploring}: ``COVID" group with positive test results within 14 days prior to the recording (141 samples), ``non-COVID without symptom" group with clean medical records (298 samples), and ``non-COVID with symptom" group with clean medical records albeit with a cough symptom (32 samples). The ``non-COVID" participants were recruited from countries where the virus was not widespread during the time of data collection, had not tested positive for COVID-19.

Each participant provided a written report of symptoms, and uploaded a recording containing five breathing or three cough samples through either an Android application or a web-based platform. Each sample corresponded to a single audio recording. Recordings were imported at a sampling rate of 22050 Hz using the Librosa library \cite{mcfee2015librosa}. Silent or noisy recordings were discarded. Silent periods at the beginning and end of the recording were trimmed. Only recordings that were longer than 2 s were analyzed. \revhl{To alleviate data imbalance, augmentation was performed on the training set via amplification by a random scalar in [1.15 2], random change of pitch speed by [0.8 0.99], and addition of white noise without distorting the audibility of the original recordings significantly, following procedures in \cite{brown2020exploring}.}

Two separate tasks were considered: discriminating ``COVID" versus ``non-COVID without symptom" groups (Task 1), and ``COVID" versus ``non-COVID with symptom" groups (Task 2). Both tasks were implemented given either cough or breath modalities as input \cite{ni2021automated}. \revhl{In Task 1, we analyzed (137, 141) samples in COVID and non-COVID groups respectively for cough modality, (141, 144) samples in COVID and non-COVID groups for breath modality. In Task 2, we analyzed (54, 88) samples in COVID versus non-COVID groups for cough modality, (56, 89) samples in COVID versus non-COVID groups for breath modality.}

 \subsubsection{COUGHVID Dataset}
\revhl{The dataset included participants of different ages, genders, geographic locations, and COVID-19 statuses \cite{orlandic2021coughvid}. Four experienced physicians labeled the recordings to diagnose any pulmonary abnormalities. Here, two groups of subjects were analyzed: ``COVID" group with disease labels and cough symptoms (608 samples), and the ``non-COVID with symptom" group with clean medical records and cough symptoms (1778 samples). Cough sounds were captured through a web-based platform. To improve data quality, poor recordings with signal-to-noise ratio (SNR) lower than 0.8 were discarded. A spectral peak detection algorithm was used to segment each recording into individual cough events. The recordings were imported using the Librosa library \cite{mcfee2015librosa} at a sampling rate of 22050 Hz. Silent periods present at the start and end of the recordings were trimmed. Only recordings that were longer than 2 s were analyzed. Data augmentation was performed on the training set to alleviate data imbalance. A single task given cough modality as input was considered where ``COVID" versus ``non-COVID with symptom" groups were discriminated. Accordingly, we analyzed (1644, 1644) samples in COVID versus non-COVID groups.}

  \begin{figure} [ht]
   \centering
   \begin{tabular}{c}
   \includegraphics[width=0.95\columnwidth]{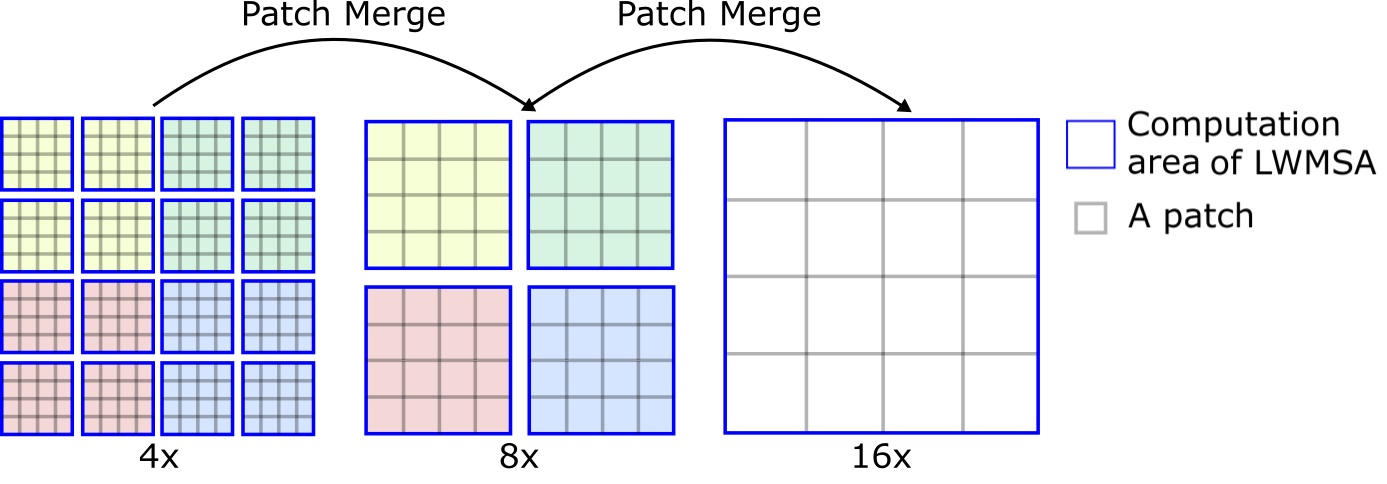}
  \end{tabular}
   \caption
   { \label{fig:lwmsa} 
Illustration of the hierarchical construct for the local windowed multi-head self-attention (LWMSA) mechanism in HST. In a given stage, attention is computed among a local neighborhood of $M \times M$ patches ($M=4$ in this example). During progression onto the next stage, a feature maps are concatenated via patch merge across $2 \times 2$ grids of neighboring patches, resulting in 4-fold increase in patch size. LWMSA is again performed in a neighborhood of $M \times M$ patches, but a broader receptive field is covered across the spectrogram due to growing patch size.}
   \end{figure}

\subsection{Hierarchical Spectrogram Transformer}
The proposed method leverages a novel architecture to detect COVID-19 from respiratory sounds. Audio recordings of cough or breathing sounds are first mapped onto a spectrogram that captures a two-dimensional time-frequency representation (Section \ref{sec:spectrogram}). The spectrogram is then processed with a hierarchical patch-based attention mechanism to capture local to global context across network layers (Section \ref{sec:network}). Here, we introduce a five-stage model where the first four stages derive latent audio representations with a cascade of local-windowed transformer blocks (Section \ref{sec:lwattention}). The spectrotemporal resolution of feature maps is progressively lowered across these stages, while the channel dimensionality is expanded. Meanwhile, the last stage maps latent representations onto a classification output via cross-entropy loss (Section \ref{sec:celoss}). \revhl{Code for implementing HST is available at \href{https://github.com/icon-lab/HST}{https://github.com/icon-lab/HST}}. 

\subsubsection{Spectrogram Features}
\label{sec:spectrogram}
Many prior studies on disease detection from audio recordings suggest that spectrotemporal features can provide comprehensive description of respiratory sounds  \cite{pramono2019evaluation,bahoura2009pattern,7280317}. Inspired by this success, here we employ \revhl{Mel-spectrogram} representations of audio recordings. Given a recording of duration D, a the short-time Fourier transform (STFT) of audio recordings are calculated over windowed segments. The STFT ($A[k, n]$) of the sampled audio signal ($a[n]$) at time $n$ with length $N$ is taken as:
\begin{equation}
A[k, n]= \sum_{m=-\infty}^{\infty}a[m] \cdot w[m-n] \cdot \mathrm{exp}\left(-\frac{j2{\pi}km}{N}\right)
\end{equation}
for $k=0,1,2,...,N-1$. The computed frequency components are $f=\frac{kf_s}{N}$ where the sampling frequency is $f_s$=22050 Hz, $N = 2048$ and $w[n]$ is the Hanning Window of length $N$.
\begin{equation}
w[n]=0.5-0.5 \cdot \mathrm{cos}\left(\frac{2{\pi}n}{N}\right)
\end{equation}
The STFT is computed via a 2048-point fast Fourier transform (FFT) with 2048 points in each segment and 128 overlapping points between consecutive segments, \revhl{for a favorable trade-off between spectral and temporal resolution}. Lastly, the magnitude square of the STFT coefficients are used to derive the spectrogram representation:
\begin{equation}
\mbox{\revhl{Mel}-spectrogram}[mel,n]=\mathrm{log}({|A[\revhl{{f(mel)N}/{f_s}},n]|}^2)
\end{equation}
\revhl{where $f$(mel) denotes the frequency in Hz corresponding to the mel-scale frequency $mel$}. Two-dimensional (2D) spectrogram coefficients are log transformed to induce compressive nonlinearity and linearly downsampled onto a 224x224 matrix for efficiency.

\subsubsection{Network Architecture}
\label{sec:network}
Here, we propose a five-stage architecture for COVID-19 detection (Fig. \ref{fig:architecture}). In Stage 1, the 2D spectrograms are partitioned into non-overlapping patches of size $h \times h$, where $h=4$. Thus, partitioning produces a grid of $P \times P =56 \times 56$ patches. Each patch is flattened onto a feature dimensionality of $4 \times 4 = 16$, and then projected with a linear embedding layer onto a dimensionality $d$, where $d=96$. This provides an input feature map of size $56\times56\times d$ to the HST block in Stage 1. The input map is then processed with the transformer block equipped with local windowed attention mechanisms to achieve linear complexity as described in the following section, instead of global attention mechanisms in vanilla transformer models that suffer from quadratic complexity \cite{vaswani2017attention,gong2021ast}. 

HST comprises a hierarchical architecture for efficient capture of contextual features in spectrograms across multiple scales. To do this, the spectrotemporal resolution of features maps is progressively decreased while the embedding dimensionality is increased across stages (Fig. \ref{fig:lwmsa}). In Stages 2-4, a patch merge layer is used across $2 \times 2$ grids of neighboring patches to lower the number of patches (i.e., sequence tokens) by a factor of 4. Afterwards, a linear embedding layer scales up the embedding dimensionality by a factor of 2. Therefore, the input feature maps to the stages are given as $(56/2^{S-2})\times(56/2^{S-2})\times (2^{S-2}d)$, and the input feature maps to the HST blocks in each stage are given as $(56/2^{S-1})\times(56/2^{S-1})\times (2^{S-1}d)$, where $S$ is the stage number (see Fig. \ref{fig:architecture}). A cascade of transformer blocks are then employed in each stage, where local windowed attention is computed over a broader scale due to merged patches. 

In its final stage, HST processes latent representations of spectrograms extracted via the prior stages for disease detection. In Stage 5, an input feature map of size $7\times7\times8d$ is received. Afterwards, this feature map is passed through a cascade of a normalization layer, a one-dimensional adaptive average pooling layer \cite{gong2021ast}, and a linear classification head with two output units.

Small, base and large variants of HST were implemented. Across Stages 1-4, the small variant had (1,1,3,1) blocks, hidden size of 96-768, MLP size of 384-3072, 3-24 attention heads, the base variant had (1,1,9,1) blocks, hidden size of 96-768, MLP size of 384-3072, 3-24 attention heads, and the large variant had (1,1,9,1) blocks, hidden size of 128-1024, MLP size of 512-4096, 4-32 attention heads.

\subsubsection{Local Windowed Transformer Blocks}
\label{sec:lwattention}
Each stage of HST except for the final stage includes a cascade of transformer blocks to derive attention-based latent representations. The transformer blocks are composed of multi-head self-attention (MSA) modules and MLPs, interleaved with normalization layers and residual connections as in \cite{vaswani2017attention}. However, unlike vanilla token-based \cite{baevski2020wav2vec} or patch-based transformers \cite{gong2021ast} proposed for auditory tasks, the MSA modules in HST leverage local windowed attention as inspired by the the success of restricted attention models in computer vision tasks \cite{yang2021focal,liu2021Swin,korkmaz2022unsupervised}. Attention is restricted to a local neighborhood of $M \times M$ patches, where $M=7$ in this work. Within a transformer block, a cyclic shift of $(\lfloor{\frac{M}{2}}\rfloor,\lfloor{\frac{M}{2}}\rfloor)$ patches is enforced to improve diversity in window definition prior to the second MSA module. Given the input feature map $y^{i-1}$ to the $i^{th}$ transformer block, latent representations are computed as:
\begin{equation} 
\label{eqs}
\begin{split}
&\hat{z}^i = \mathrm{LWMSA}(\mathrm{LN}(y^{i-1}))+y^{i-1}, \\
&z^i = \mathrm{MLP}(\mathrm{LN}(\hat{z}^i))+\hat{z}^i,\\
&\hat{y}^{i} = \mathrm{LWMSA}(\mathrm{LN}(z^{i}))+z^{i},\\
&y^{i} = \mathrm{MLP}(\mathrm{LN}(\hat{y}^{i}))+\hat{y}^{i},
\end{split}
\end{equation}
where LWMSA is the local windowed MSA module, LN denotes layer normalization, ${y}^{l}$ is the output of the transformer block. A two-layer MLP is used with Gaussian error linear unit (GELU) activation functions. 

Provided an input sequence of $n_t$ tokens $X\in \mathbb{R}^{n_t \times d}$, the self-attention matrix in an LWMSA module is calculated as: 
\begin{equation}
\label{eq:attention}
\mathrm{Attention}(Q, K, V) = \mathrm{softmax}\left(\frac{QK^T}{\sqrt{d}}+B\right)V,
\end{equation}
where $B\in \mathbb{R}^{n_t \times n_t}$ is a relative position bias matrix \cite{vaswani2017attention}, and $Q, K, V\in \mathbb{R}^{n_T \times d}$ are the query, key, and value matrices obtained as learnable linear projections of $X$:
\begin{equation}
\begin{split}
Q=XL^Q, \\
K=XL^K,\\
V=XL^V,\\
\end{split}
\end{equation}
where $L^Q,L^K,L^V \in \mathbb{R}^{n_t \times d}$ are the corresponding projection matrices. Note that, in Eq. \ref{eq:attention}, calculation of softmax attention involves formation of an $\mathbb{R}^{n_t \times n_t}$ inter-token interaction matrix. A single sequence with $n_t=P^2$ patches are processed in vanilla MSA modules \cite{vaswani2017attention,dalmaz2021resvit}, resulting in a quadratic complexity of $O(P^2 \times P^2)$. In contrast, LWMSA splits the overall sequence into $(\frac{P}{M})^2$ sub-sequences of length $n_t=M^2$ each. As such, LWMSA enables a linearly scaled complexity of $O((\frac{P}{M})^2 \times M^2 \times M^2)=O(P^2 \times M^2)$ with respect to sequence length.  

  \begin{figure} [t]
   \centering
   \begin{tabular}{c}
   \includegraphics[width=0.85\columnwidth]{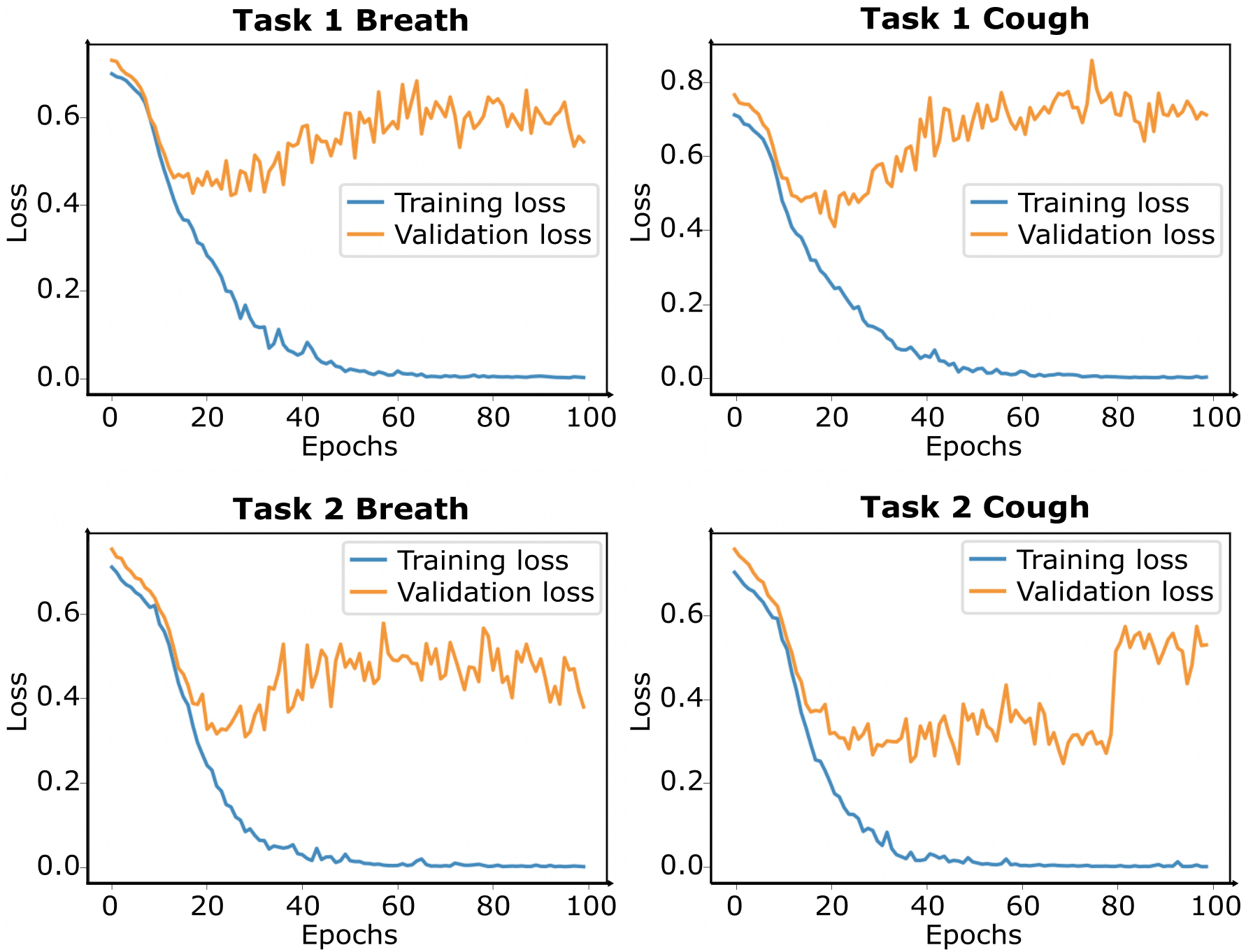}
  \end{tabular}
   \caption
   { \label{fig:loss} 
\revhl{Training and validation losses of HST across training epochs. Results shown separately for models based on breath and cough modalities, and for Task 1 (COVID vs. non-COVID without symptoms) and Task 2 (COVID vs. non-COVID with symptoms).}}
   \end{figure}
   
\begin{table}[t]
\caption{Validation F1 scores of competing neural network models across learning rates. Highest performing rates for each model are marked in \textbf{bold} font.} 
\label{tab:lrtuning}
\centering 
\resizebox{0.85\columnwidth}{!}{%
\begin{tabular}{lccccc}
\toprule 
&  $10^{-5}$    & $3 \times 10^{-4}$ & $10^{-3}$ & $10^{-2}$  \\
\midrule 
CNN & 0.77$\pm$0.10 & 0.79$\pm$0.10 & \textbf{0.82}$\pm$\textbf{0.05} & 0.75$\pm$0.13 \\
DeepShufNet & 0.68$\pm$0.11 & 0.71$\pm$0.13 & 0.67$\pm$0.15 & \textbf{0.77}$\pm$\textbf{0.10}  \\
LSTM & 0.52$\pm$0.17 & 0.55$\pm$0.24 & \textbf{0.56}$\pm$\textbf{0.10} & 0.54$\pm$0.25  \\
BiLSTM & 0.47$\pm$0.08 & 0.58$\pm$0.07 & \textbf{0.60}$\pm$\textbf{0.08} & 0.57$\pm$0.08  \\
CNN-LSTM & 0.51$\pm$0.18 & \textbf{0.69}$\pm$\textbf{0.09} & 0.52$\pm$0.19 & 0.53$\pm$0.14  \\
A-CNN-LSTM & 0.54$\pm$0.28 & 0.68$\pm$0.08 & 0.71$\pm$0.10 & \textbf{0.73$\pm$0.12 } \\
Wav2Vec & \textbf{0.67}$\pm$\textbf{0.04} & 0.61$\pm$0.05 &  0.57$\pm$0.03 & 0.58$\pm$0.03  \\
AST & 0.71$\pm$0.12 & 0.69$\pm$0.11 & \textbf{0.74}$\pm$\textbf{0.14} & 0.72$\pm$0.15  \\
HMT & \textbf{0.77}$\pm$\textbf{0.09} & 0.76$\pm$0.08 & 0.73$\pm$0.06 & 0.67$\pm$0.24  \\
Ensemble & 0.70$\pm$0.16 & 0.75$\pm$0.13 & \textbf{0.77}$\pm$\textbf{0.16} & 0.70$\pm$0.10  \\
HST & \textbf{0.83}$\pm$\textbf{0.07} & 0.74$\pm$0.08 & 0.70$\pm$0.10 & 0.72$\pm$0.15  \\
\bottomrule 
\end{tabular}}
\end{table}

\begin{table}[t]
\caption{Hyperparameters for neural network models.} 
\label{tab:hype}
\begin{center}
\resizebox{0.8\columnwidth}{!}{%
\begin{tabular}{lcccc}
\hline
   & Learning Rate & $\lambda_{L_2}$ & Batch Size & Optimizer\\
\hline
CNN &  $10^{-3}$ & $10^{-4}$ & $8$ & Adam \\
DeepShufNet &  $10^{-2}$ & $10^{-4}$ & $8$ & Adam  \\
LSTM &  $10^{-3}$ & - & $8$ & Adam \\
BiLSTM &  $10^{-3}$ & - & $8$ & Adam \\
CNN-LSTM &  $3 \times 10^{-4}$ & - & $8$ & Adam\\
A-CNN-LSTM &  $ 10^{-2}$ & $10^{-4}$ & $8$ & Adam\\
Wav2Vec &  $10^{-5}$ & $10^{-6}$ & $32$ & Adam \\
AST &  $10^{-3}$ & $10^{-4}$ & $8$ & Adam  \\
HMT &  $10^{-5}$ & $10^{-4}$ & $5$ & Adam  \\
Ensemble &  $10^{-3}$ & $10^{-4}$ & $6$ & Adam \\
HST &  $10^{-5}$ & $10^{-8}$ & $8$ & AdamW \\
\hline
\end{tabular}
}
\end{center}
\end{table}

\subsubsection{Loss Function}
\label{sec:celoss}
A binary cross-entropy loss function is employed to train the HST models for COVID-19 detection. The classification head at the output layer produces a predicted probability for each output class in the range [0, 1]. Cross-entropy loss is then expressed as:
\begin{equation}
\mathrm{Loss} = -\frac{1}{J}\sum_{j=1}^{J}(x_j \mathrm{log}(\hat{x}_j)+(1-x_j) \mathrm{log}(1-\hat{x}_j))
\end{equation}
where $x_j$ is the true label of the $j^{th}$ sample (1 denoting COVID-19, 0 denoting healthy control), $\hat{x}_j$ is the predicted probability of the $j^{th}$ sample, and $J$ is the number of samples.

\begin{table}[t]
\caption{\revhl{Test performance of competing methods in Task 1 based on cough sounds for the Cambridge dataset. Performance in distinguishing the COVID from the non-COVID group without symptoms is listed as mean$\pm$std across cross-validation folds.}} 
\label{tab:task1cough}
\centering
\resizebox{0.85\columnwidth}{!}{%
\begin{tabular}{lccccc}
\toprule
 & AUC & Precision & Recall & F1 \\
\midrule
SVM & 0.68$\pm$0.13 & 0.64$\pm$0.12 & 0.72$\pm$0.13 & 0.67$\pm$0.10   \\
CNN & 0.93$\pm$0.04 & 0.87$\pm$0.10 & 0.86$\pm$0.08 & 0.86$\pm$0.06  \\
DeepShufNet & 0.92$\pm$0.04 & 0.81$\pm$0.11 & 0.81$\pm$0.08 & 0.80$\pm$0.05  \\
LSTM & 0.71$\pm$0.10 & 0.64$\pm$0.07 & 0.68$\pm$0.09 & 0.66$\pm$0.07  \\
BiLSTM & 0.64$\pm$0.08 & 0.62$\pm$0.08 & 0.68$\pm$0.16 & 0.64$\pm$0.09  \\
CNN-LSTM & 0.66$\pm$0.12 & 0.64$\pm$0.09 & 0.61$\pm$0.19 & 0.61$\pm$0.10  \\
A-CNN-LSTM & 0.88$\pm$0.05 & 0.81$\pm$0.11 & 0.71$\pm$0.09 & 0.75$\pm$0.08  \\
Wav2Vec & 0.82$\pm$0.04 & 0.83$\pm$0.06 & 0.70$\pm$0.03 & 0.76$\pm$0.03  \\
AST & 0.86$\pm$0.04 & 0.89$\pm$0.11 & 0.70$\pm$0.13 & 0.77$\pm$0.05  \\
HMT & 0.89$\pm$0.07 & 0.77$\pm$0.11 & 0.77$\pm$0.13 & 0.76$\pm$0.05  \\
Ensemble & 0.90$\pm$0.04 & 0.84$\pm$0.12 & 0.74$\pm$0.16 & 0.77$\pm$0.08  \\
HST & \textbf{0.97}$\pm$\textbf{0.03} & \textbf{0.92}$\pm$\textbf{0.08} & \textbf{0.94}$\pm$\textbf{0.07} & \textbf{0.93}$\pm$\textbf{0.04}  \\
\bottomrule
\end{tabular}}
\end{table} 

\subsection{Competing Methods}\label{sec:Classifiers}

Several state-of-the-art baselines were adopted including a traditional ML method along with CNN, RNN, transformer, and ensemble models for COVID-19 detection. \revhl{All models analyzed grayscale spectrograms, yet the spectrogram for a given audio recording was replicated across the three color channels to provide inputs to CNN modules}. 

\textbf{SVM:} A total of 733 hand-crafted and data-driven features were taken as inputs \cite{brown2020exploring}. Hand-crafted features including duration, onset, tempo, period, RMS energy, spectral centroid, roll-off frequency, zero crossing, MFCC, $\Delta$-MFCC, $\Delta^{2}$-MFCC \revhl{were extracted via the Librosa library \cite{mcfee2015librosa}}. Data-driven features were obtained from intermediate layers of a pre-trained \emph{VGGish} model. An SVM classifier with a radial basis function (RBF) kernel was built for each task.

\textbf{CNN:} A CNN model was built that received as input spectrogram features \cite{bagad2020cough}. The ResNet34 architecture was adopted with image resolution 224x224, albeit a fully-connected (FC) layer with two output units was used for COVID-19 detection.

\revhl{\textbf{DeepShufNet:} A CNN model was built based on the efficient architecture reported in \cite{deepshufnet}. The models received as input spectrogram features identical to the CNN baseline.}

\textbf{LSTM:} An RNN model was built based on the LSTM architecture in \cite{pahar2021covid}. Model inputs included \revhl{13 MFCC, 13 $\Delta$-MFCC, 13 $\Delta^{2}$-MFCC}, 1 spectral center, 7 spectral contrast and 12 chroma features. A total of 59 features were compiled. 

\textbf{BiLSTM:} A bidirectional RNN model was built based on the BiLSTM architecture reported in \cite{balamurali2021deep}. The input features were identical to those for the LSTM baseline. 

\textbf{CNN-LSTM:} A hybrid model composed of CNN and LSTM layers was adapted as reported in \cite{islam2020combined}. The input features matched those provided to the LSTM baseline. 

\revhl{\textbf{A-CNN-LSTM:} An attention-based hybrid CNN-LSTM was implemented as reported in \cite{hamdi2022cnnlstm}. The model received as input spectrogram features.}

\textbf{Wav2Vec:} An audio transformer model was adopted \cite{baevski2020wav2vec}. Wav2Vec processed raw audio signals resampled at 16 KHz with a feature encoder (7-layer CNN), a context encoder (12 transformer blocks), followed by a classification layer.

\textbf{AST:} An audio spectrogram transformer (AST) introduced for general audio tasks was adopted \cite{gong2021ast}. The model received as input spectrogram features as in HST. The architecture comprised vanilla transformer blocks processing the input with patch size 16x16. The last layer of AST was adapted to a classification layer with two output units.

\revhl{\textbf{HMT:} A hierarchical multi-modal transformer (HMT) for COVID-19 detection was implemented \cite{hmt}. The model fused representations from a transformer branch processing spectrograms and from an MLP branch processing MFCC features.}

\textbf{Ensemble:} An ensemble model \cite{9362499} was implemented that combined CNN and AST models described above. Output feature maps prior to classification layers in each model were concatenated, and input to an FC layer with two output units.

\subsection{Modeling Procedures}
\revhl{Models were implemented in PyTorch except for} LSTM, BiLSTM and CNN-LSTM that were implemented in Tensorflow. Models were executed on NVidia A4000 GPUs. Spectrogram intensities were normalized to a mean of 0.5 and standard deviation of 0.5. \revhl{Modeling was performed separately on each dataset via 10-fold cross validation} \cite{brown2020exploring}, with data partitioned in each fold into non-overlapping training, test and validation sets (70\%, 20\%, 10\%). Model parameters were estimated on the training set. Following common practice in literature, transformer models were initiated with pre-trained weights. AST and HST were pre-trained for object detection on ImageNet, Wav2Vec was pre-trained on sampled speech audio from the LibriSpeech dataset. \revhl{Training was continued for a maximum of 100 epochs (see Fig. \ref{fig:loss} for HST loss across epochs).} Gradient clipping was used with an upper threshold of 0.1 for the gradient norm.

\begin{table}[t]
\caption{Performance in Task 1 based on breathing sounds, distinguishing the COVID from the non-COVID group without symptoms.} 
\label{tab:task1breath}
\centering
\resizebox{0.85\columnwidth}{!}{%
\begin{tabular}{lccccc}
\toprule
 & AUC & Precision & Recall & F1\\
\midrule
SVM & 0.65$\pm$0.14 & 0.63$\pm$0.14 & 0.70$\pm$0.19 & 0.66$\pm$0.15  \\
CNN & 0.91$\pm$0.04 & 0.82$\pm$0.12 & 0.72$\pm$0.18 & 0.74$\pm$0.09  \\
DeepShufNet & 0.91$\pm$0.07 & 0.83$\pm$0.07 & 0.85$\pm$0.07 & 0.84$\pm$0.06  \\
LSTM & 0.68$\pm$0.14 & 0.65$\pm$0.19 & 0.60$\pm$0.16 & 0.62$\pm$0.16  \\
BiLSTM & 0.69$\pm$0.12 & 0.63$\pm$0.10 & 0.68$\pm$0.19 & 0.65$\pm$0.14  \\
CNN-LSTM & 0.72$\pm$0.09 & 0.68$\pm$0.08 & 0.75$\pm$0.13 & 0.71$\pm$0.08  \\
A-CNN-LSTM & 0.84$\pm$0.06 & 0.83$\pm$0.11 & 0.63$\pm$0.14 & 0.71$\pm$0.11  \\
Wav2Vec & 0.83$\pm$0.04 & 0.81$\pm$0.05 & 0.67$\pm$0.08 & 0.73$\pm$0.06  \\
AST & 0.85$\pm$0.01 & 0.82$\pm$0.10 & 0.70$\pm$0.19 & 0.75$\pm$0.13  \\
HMT & 0.86$\pm$0.09 & 0.82$\pm$0.09 & 0.71$\pm$0.17 & 0.75$\pm$0.13  \\
Ensemble & 0.87$\pm$0.05 & 0.80$\pm$0.06 & 0.80$\pm$0.07 & 0.80$\pm$0.04  \\
HST & \textbf{0.97}$\pm$\textbf{0.02} & \textbf{0.94}$\pm$\textbf{0.06} & \textbf{0.95}$\pm$\textbf{0.04} & \textbf{0.94}$\pm$\textbf{0.04}  \\
\bottomrule
\end{tabular}
}
\end{table}

\begin{figure}[t]
   \centering
   \begin{tabular}{c}
   \includegraphics[width=0.7\columnwidth]{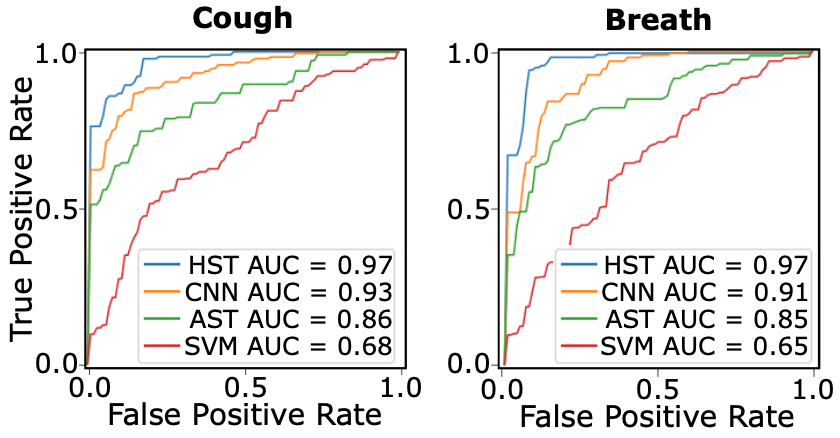}
   \end{tabular}
    \caption{ROC curves for Task 1 to distinguish the COVID group from the non-COVID group without symptoms, based on cough (left), or breathing sounds (right). AUC is listed for each model (see legend).}
    \label{fig:roccurves1}
\end{figure}

Model hyperparameters were selected based on F1 performance on the validation set (see Table \ref{tab:lrtuning} for representative results on learning rate, and Table ~\ref{tab:hype} for selected parameters). For SVM, regularization parameter \emph{C} and kernel coefficient $\gamma$ were selected via grid search. For network models, a common set of learning rate, regularization parameter for $L_2$ norm of model weights ($\lambda_{L_2}$), batch size, and optimizer were selected across tasks. \revhl{The number of epochs was selected separately for each task.} Model performance was measured using AUC, precision, recall, and F1 metrics on the test set. Individual metrics were averaged across the test set. \revhl{Mean and standard deviation of metrics were then reported across cross-validation folds. Statistical significance of performance differences was assessed via non-parametric Wilcoxon signed-rank tests.}

\section{Results}
\label{sec:sections}

\begin{table}[t]
\caption{Performance in Task 2 based on cough sounds, distinguishing the COVID group from the non-COVID group with symptoms. } 
\label{tab:task2cough}
\centering  
\resizebox{0.85\columnwidth}{!}{%
\begin{tabular}{lccccc}
\toprule 
 & AUC & Precision & Recall & F1 \\
\midrule
SVM & 0.69$\pm$0.15 & 0.62$\pm$0.16 & 0.65$\pm$0.17 & 0.62$\pm$0.15   \\
CNN & 0.95$\pm$0.05 & 0.87$\pm$0.12 & 0.92$\pm$0.10 & 0.89$\pm$0.09  \\
DeepShufNet & 0.96$\pm$0.03 & 0.86$\pm$0.09 & \textbf{0.98}$\pm$\textbf{0.04} & 0.91$\pm$0.05  \\
LSTM & 0.57$\pm$0.20 & 0.56$\pm$0.18 & 0.71$\pm$0.27 & 0.62$\pm$0.20  \\
BiLSTM & 0.38$\pm$0.23 & 0.69$\pm$0.31 & 0.43$\pm$0.20 & 0.49$\pm$0.20  \\
CNN-LSTM & 0.55$\pm$0.21 & 0.67$\pm$0.17 & 0.43$\pm$0.12 & 0.49$\pm$0.13  \\
A-CNN-LSTM & 0.88$\pm$0.08 & 0.80$\pm$0.11 & 0.88$\pm$0.12 & 0.83$\pm$0.09  \\
Wav2Vec & 0.89$\pm$0.05 & 0.76$\pm$0.12 & 0.80$\pm$0.08 & 0.77$\pm$0.08  \\
AST & 0.83$\pm$0.11 & 0.86$\pm$0.06 & 0.86$\pm$0.07 & 0.86$\pm$0.05  \\
HMT & 0.87$\pm$0.07 & 0.80$\pm$0.09 & 0.85$\pm$0.07 & 0.82$\pm$0.07  \\
Ensemble & 0.94$\pm$0.06 & 0.78$\pm$0.15 & 0.96$\pm$0.09 & 0.85$\pm$0.09  \\
HST & \textbf{0.98}$\pm$\textbf{0.03} & \textbf{0.94}$\pm$\textbf{0.08} & 0.93$\pm$0.07 & \textbf{0.94}$\pm$\textbf{0.07}  \\
\bottomrule
\end{tabular}}
\end{table}

\begin{table}[t]
\caption{Performance in Task 2 based on breathing sounds, distinguishing the COVID from the non-COVID group with symptoms. } 
\label{tab:task2breath}
\centering  
\resizebox{0.85\columnwidth}{!}{%
\begin{tabular}{lccccc}
\toprule 
 & AUC & Precision & Recall & F1 \\
\midrule
SVM & 0.71$\pm$0.18 & 0.74$\pm$0.20 & 0.58$\pm$0.19 & 0.62$\pm$0.13   \\
CNN & 0.92$\pm$0.06 & 0.85$\pm$0.08 & 0.91$\pm$0.09 & 0.87$\pm$0.06  \\
DeepShufNet & 0.96$\pm$0.04 & 0.88$\pm$0.06 & 0.95$\pm$0.06 & 0.91$\pm$0.05  \\
LSTM & 0.64$\pm$0.17 & 0.45$\pm$0.16 & 0.44$\pm$0.19 & 0.44$\pm$0.17  \\
BiLSTM & 0.61$\pm$0.21 & 0.56$\pm$0.33 & 0.48$\pm$0.32 & 0.49$\pm$0.28  \\
CNN-LSTM & 0.75$\pm$0.16 & 0.49$\pm$0.19 & 0.53$\pm$0.30 & 0.49$\pm$0.21  \\
A-CNN-LSTM & 0.87$\pm$0.08 & 0.80$\pm$0.11 & 0.92$\pm$0.06 & 0.85$\pm$0.06  \\
Wav2Vec & 0.92$\pm$0.05 & 0.80$\pm$0.17 & 0.88$\pm$0.04 & 0.83$\pm$0.09  \\
AST & 0.93$\pm$0.05 & 0.87$\pm$0.04 & 0.94$\pm$0.06 & 0.90$\pm$0.03  \\
HMT & 0.86$\pm$0.06 & 0.78$\pm$0.09 & 0.88$\pm$0.09 & 0.83$\pm$0.06  \\
Ensemble & 0.91$\pm$0.04 & 0.86$\pm$0.10 & 0.90$\pm$0.12 & 0.87$\pm$0.06  \\
HST & \textbf{0.97}$\pm$\textbf{0.02} & \textbf{0.91}$\pm$\textbf{0.06} & \textbf{0.96}$\pm$\textbf{0.09} & \textbf{0.93}$\pm$\textbf{0.04}  \\
\bottomrule
\end{tabular}}
\end{table}

\begin{figure}[t]
    \centering
   \begin{tabular}{c}
    \includegraphics[width=0.7\columnwidth]{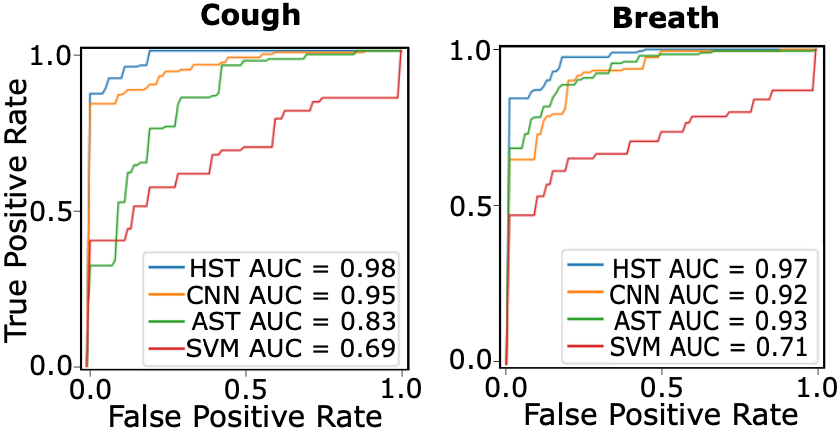}
    \end{tabular}
    \caption{ROC curves for Task 2 to distinguish the COVID-19 group from the non-COVID group with cough symptoms, based on cough sounds (left), and breathing sounds (right). AUC is listed for each model (see legend).}
    \label{fig:roccurves2} 
\end{figure}

\subsection{COVID-19 Detection}
\revhl{We first demonstrated HST on the Cambridge dataset against state-of-the-art baselines including traditional (SVM), CNN (CNN, DeepShufNet), RNN (LSTM, BiLSTM), hybrid CNN-RNN (CNN-LSTM, A-CNN-LSTM), transformer (Wav2Vec, AST, HMT), and ensemble (Ensemble) methods. In Task 1, the COVID group was distinguished from the non-COVID group without symptoms based on either cough or breathing sounds.} Performance for competing methods are reported in Table ~\ref{tab:task1cough} with cough modality, in Table ~\ref{tab:task1breath} with breathing modality. Receiver operating characteristics (ROC) curves for SVM, AST, CNN and HST are displayed in Fig. \ref{fig:roccurves1}. Overall, HST significantly outperforms competing methods ($p<0.05$). As seen in ROC curves, HST also maintains a more favorable trade-off between true and false positive rates. These results indicate that HST enables improved capture of contextual features in spectrograms to improve detection of COVID-19. Note that comparable detection performance is attained via HST with cough versus breathing modalities. \revhl{This finding implies that both modalities carry discriminative information regarding respiratory symptoms of COVID-19.} \revhl{In Task 2, the COVID group was distinguished from the non-COVID group with symptoms. Performance metrics are listed in Table \ref{tab:task2cough} for cough modality, and in Table \ref{tab:task2breath} for breathing modality. ROC curves are displayed in Fig. \ref{fig:roccurves2}. HST significantly outperforms competing methods ($p<0.05$), except for DeepShufNet and Ensemble that yield higher recall.} HST also maintains a modestly better trade-off between true and false positive rates. Taken together, these results indicate that contextual representations of audio spectrograms captured by HST enable discrimination of respiratory symptoms in COVID-19 patients versus healthy controls. 

\revhl{Next, we demonstrated HST on the separate COUGHVID dataset against the same set of traditional, CNN, RNN, hybrid CNN-RNN, transformer, and ensemble methods. A single task was implemented to distinguish the COVID group and non-COVID group with cough symptoms, based on cough sounds. Performance metrics for competing methods are reported in Table ~\ref{tab:coughvid}. All methods yield relatively lower performance on the COUGHVID dataset, implying elevated task difficulty compared to the Cambridge dataset. That said, HST again outperforms competing methods significantly in all metrics ($p<0.05$), except for DeepShufNet, A-CNN-LSTM, AST, and HMT that yield similar recall. These results suggest that HST offers more reliable capture of COVID-related cues in respiratory sounds against competing methods.}

\begin{table}[t]
\caption{\revhl{Performance in distinguishing the COVID from the non-COVID group based on cough sounds for the COUGHVID dataset.}} 
\label{tab:coughvid}
\centering
\resizebox{0.85\columnwidth}{!}{%
\begin{tabular}{lccccc}
\toprule
 & AUC & Precision & Recall & F1 \\
\midrule
SVM & 0.57$\pm$0.01 & 0.54$\pm$0.01 & 0.56$\pm$0.01 & 0.55$\pm$0.02   \\
CNN & 0.66$\pm$0.04 & 0.74$\pm$0.02 & 0.90$\pm$0.07 & 0.81$\pm$0.02  \\
DeepShufNet & 0.67$\pm$0.04 & 0.74$\pm$0.03 & \textbf{1.00}$\pm$\textbf{0.00} & 0.85$\pm$0.02  \\
LSTM & 0.55$\pm$0.01 & 0.53$\pm$0.01 & 0.64$\pm$0.07 & 0.58$\pm$0.03  \\
BiLSTM & 0.66$\pm$0.07 & 0.61$\pm$0.06 & 0.69$\pm$0.07 & 0.63$\pm$0.08  \\
CNN-LSTM & 0.71$\pm$0.03 & 0.66$\pm$0.03 & 0.76$\pm$0.09 & 0.70$\pm$0.03  \\
A-CNN-LSTM & 0.67$\pm$0.05 & 0.74$\pm$0.03 & \textbf{1.00}$\pm$\textbf{0.00} & 0.84$\pm$0.02  \\
Wav2Vec & 0.65$\pm$0.07 & 0.57$\pm$0.07 & 0.74$\pm$0.15 & 0.63$\pm$0.08  \\
AST & 0.65$\pm$0.03 & 0.73$\pm$0.02 & \textbf{1.00}$\pm$\textbf{0.00} & 0.84$\pm$0.01  \\
HMT & 0.71$\pm$0.05 & 0.72$\pm$0.06 & \textbf{1.00}$\pm$\textbf{0.00} & 0.84$\pm$0.04  \\
Ensemble & 0.76$\pm$0.06 & 0.72$\pm$0.10 & 0.98$\pm$0.05 & 0.82$\pm$0.07  \\
HST & \textbf{0.83}$\pm$\textbf{0.06} & \textbf{0.77}$\pm$\textbf{0.02} & \textbf{1.00}$\pm$\textbf{0.00} & \textbf{0.86}$\pm$\textbf{0.01}  \\

\bottomrule
\end{tabular}}
\end{table}

\begin{figure}[t]
   \begin{center}
   \begin{tabular}{c}
    \includegraphics[width=0.85\columnwidth]{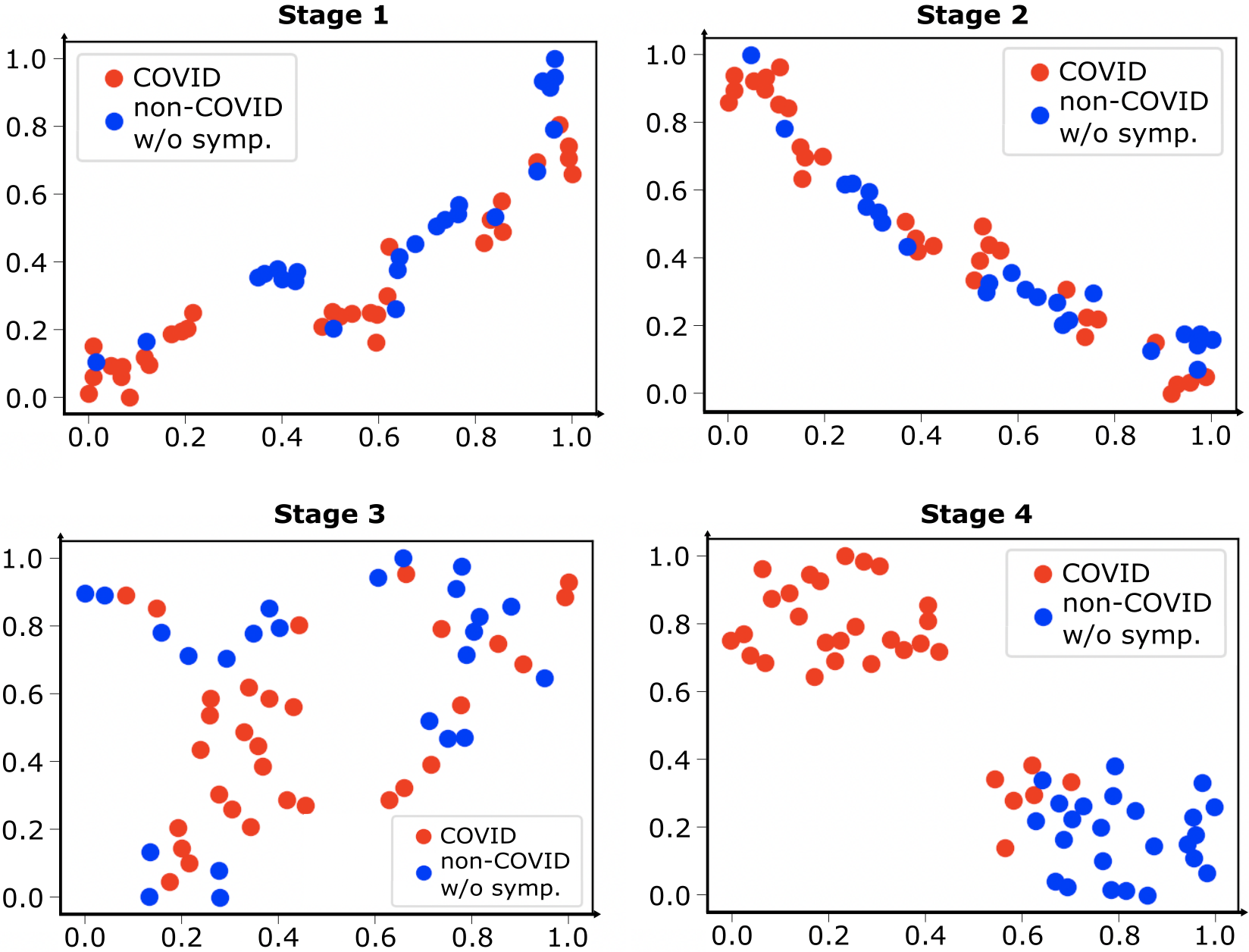}
   \end{tabular}
   \end{center}
   \caption
   { \label{fig:tsne} 
\revhl{Visualization of latent representations captured by HST across Stages 1-4. Embeddings of samples from the COVID group (red) and the non-COVID group without symptoms (blue) are displayed.}}
\end{figure} 
   
\begin{figure}[ht]
   \begin{center}
   \begin{tabular}{c} 
   \includegraphics[width=0.75\columnwidth]{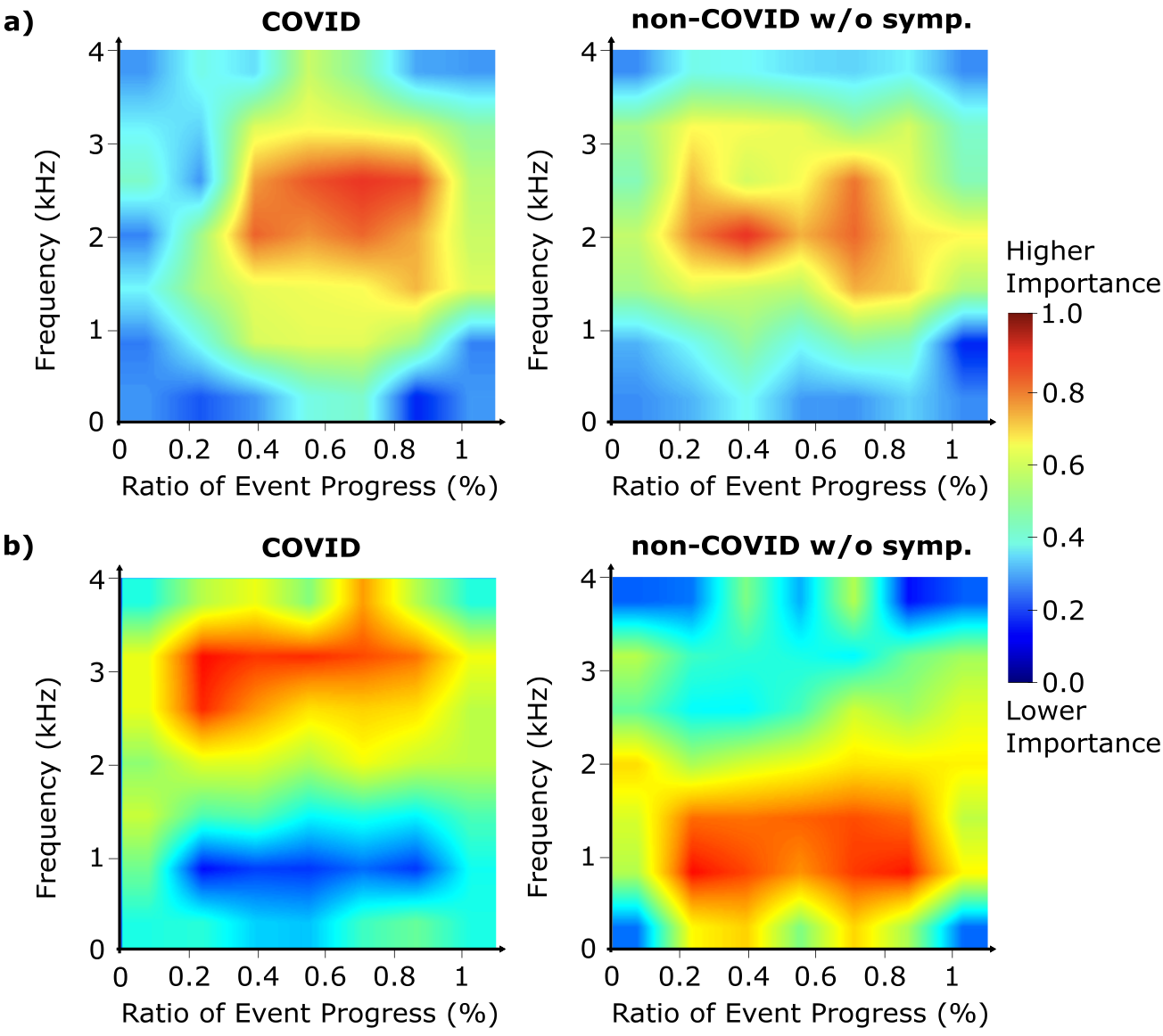}
   \end{tabular}
   \end{center}
   \caption
   { \label{fig:task1pca} 
Activation maps in HST for interpreting the spectrogram features that contribute to detection of the COVID group and the non-COVID group without symptoms. The first PC of activation maps prior to the final attention layer were computed. \revhl{Horizontal axis shows percentage event progress, i.e. the ratio of the time reached to the total duration of the recording.}
Results are shown for models based on (a) cough sounds, (b) breathing sounds in Task 1.}
\end{figure}

\begin{figure}[t]
   \begin{center}
   \begin{tabular}{c} 
   \includegraphics[width=0.75\columnwidth]{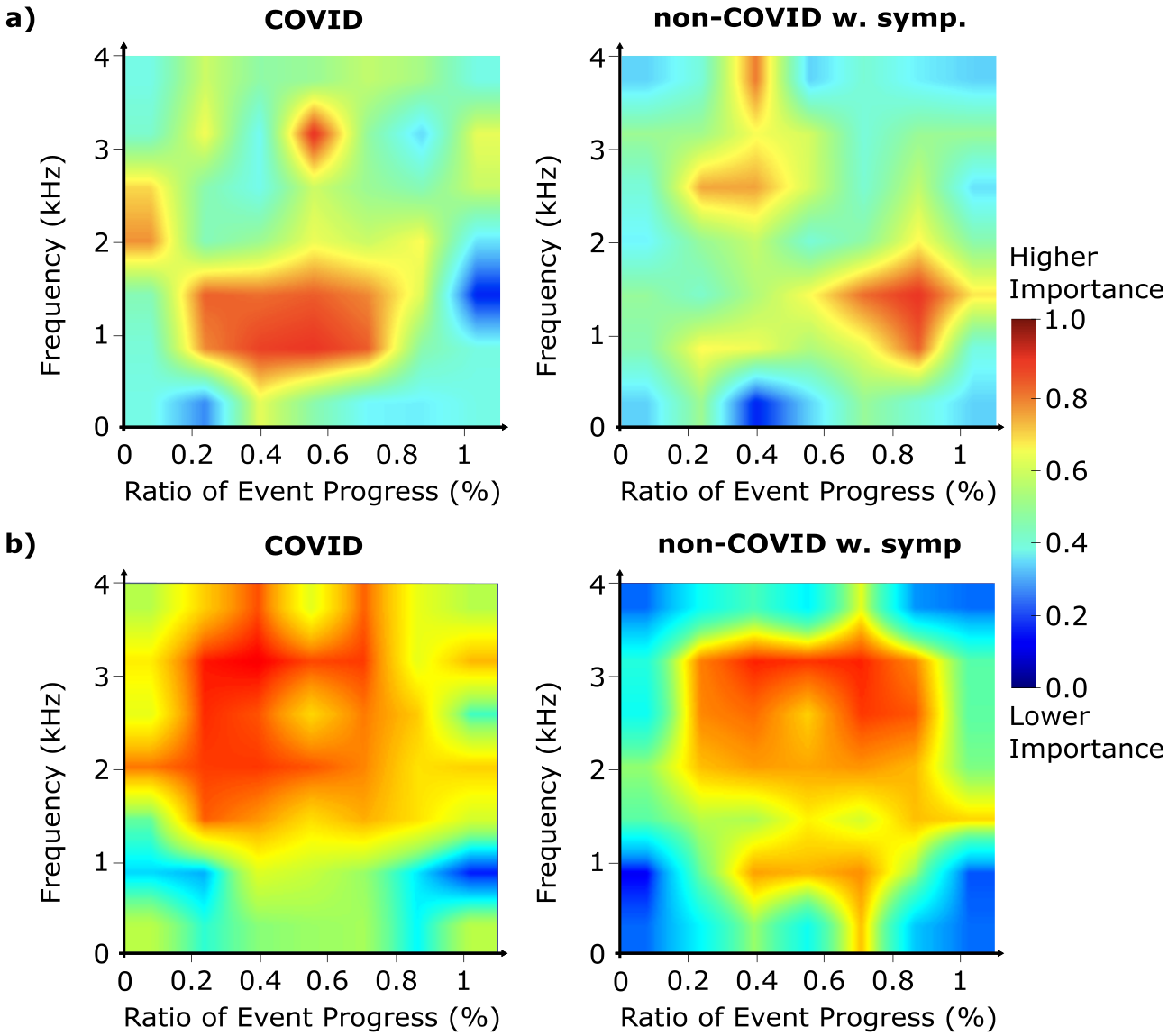}
   \end{tabular}
   \end{center}
   \caption
   { \label{fig:task2pca} 
Activation maps in HST for interpreting the spectrogram features that contribute to detection of the COVID group and the non-COVID group with symptoms. Results are shown for models based on (a) cough sounds, (b) breathing sounds in Task 2.}
   \end{figure}

\subsection{Model Interpretation}
\revhl{Across multiple stages, the proposed HST model nonlinearly transforms spectrogram features of audio recordings to extract their latent representations. Ideally, these representations should capture hidden relationships among input features that serve to improve discrimination between output classes. To examine this issue, we visualized the latent representations extracted across separate stages in HST. A random subset of 55 audio recordings were projected through HST, and the evoked hidden unit responses for each sample were stored as stage-specific response vectors. At each stage, the response vectors of all samples were embedded via t-SNE onto two dimensions \cite{tsne}. Figure \ref{fig:tsne} displays representative embeddings of samples from the COVID and non-COVID without symptoms groups based on breath modality in the Cambridge dataset. The latent representations of COVID and non-COVID samples become progressively more distinct across stages, indicating that hierarchical transformations in HST extract latent time-frequency features critical in disease detection.}

Next, a post-hoc explanatory analysis was performed to interpret the spectrogram features that most significantly contribute to model decisions. This analysis was conducted via the Grad-CAM method on correctly classified test samples from the COVID and non-COVID groups in the Cambridge dataset. \revhl{For a given sample, Grad-CAM leverages the model gradients to produce activation maps that help localize the input features predominantly contributing to the model's decision \cite{selvaraju2017grad}. For HST, these activation maps indicate the relative importance levels of spectrogram features for detecting the COVID and non-COVID groups. To obtain characteristic maps for each group, principal component analysis (PCA) was performed on activation maps across samples within each group.} The first PCs are shown in Fig.~\ref{fig:task1pca} for Task 1, and in  Fig.~\ref{fig:task2pca} for Task 2. For tasks implemented on the cough modality, the frequency bands attended by HST are roughly similar between COVID patients and healthy controls. Yet, the COVID group shows uniform emphasis on the frequency band persistently across the audio recording, whereas the non-COVID groups show relatively heterogeneous emphasis across frequencies and time. For tasks implemented on the breathing modality, the COVID group shows elevated emphasis on higher frequencies (centered around \revhl{2-3} kHz), whereas the non-COVID group has emphasis on relatively lower frequencies (as low as \revhl{1-2} kHz). Note that the differences between activation maps for COVID versus non-COVID groups are more accentuated for Task 1, albeit modest for Task 2. This can be attributed to the presence of cough as a common symptom across groups in Task 2. Although clinical research regarding the frequency features of respiratory sounds in COVID-19 is ongoing, a prior study has examined respiratory symptoms of COVID-19 patients in different stages of disease \cite{davidsonacoustic}. Frequency of respiration sounds was reported to be increase with disease development. While our results are consistent with this finding, future work on larger patient cohorts is warranted to examine the validity of the observations reported here.

\subsection{Ablation Studies}
Ablation studies were conducted to examine the influences of \revhl{input spectrogram type, spectrogram window length, model size, and pre-training on the test performance of HST on the Cambridge dataset.} The results for all four performance metrics (i.e., AUC, precision, recall, F1) are consistent, albeit only F1 scores for the variant models are listed for brevity. \revhl{Table \ref{tab:spectuning} lists performance for the proposed Mel-spectrogram and alternative CQT, Gammatone, Bark and Linear spectrograms. Table \ref{tab:lengthtuning} lists performance for the proposed 2048 window length and alternative 1024 and 4096 lengths. Table \ref{tab:sizetuning} lists performance for the proposed base model size and alternative small and large sizes. Table \ref{tab:traintuning} lists performance with and without pre-training, where HST is pre-trained for object classification on natural images. For all tasks and modalities, the proposed configuration of the HST model yields higher performance in all four tasks against variant models.} 

\begin{table}[t]
\caption{\revhl{Test F1 scores of HST in Tasks 1 and 2 based on cough or breathing sounds with  different time-frequency representations.}} 
\label{tab:spectuning}
\centering 
\resizebox{\columnwidth}{!}{%
\begin{tabular}{lccccc}
\toprule 
&  Mel & CQT & Gammatone & Bark & Linear \\
\midrule
Task 1 Cough & \textbf{0.93}$\pm$\textbf{0.04} & 0.84$\pm$0.06 & 0.87$\pm$0.07 & 0.85$\pm$0.03 & 0.86$\pm$0.06  \\
Task 1 Breath & \textbf{0.94}$\pm$\textbf{0.04} & 0.85$\pm$0.03 & 0.83$\pm$0.07 & 0.86$\pm$0.03 & 0.86$\pm$0.02  \\
Task 2 Cough & \textbf{0.94}$\pm$\textbf{0.07} & 0.92$\pm$0.07 & 0.93$\pm$0.02 & 0.92$\pm$0.01 & 0.92$\pm$0.02  \\
Task 2 Breath & \textbf{0.93}$\pm$\textbf{0.04} & 0.90$\pm$0.03 & 0.90$\pm$0.06 & 0.91$\pm$0.04 & 0.90$\pm$0.08  \\
\bottomrule 
\end{tabular}}
\end{table}

\begin{table}[t]
\caption{\revhl{Test F1 scores of HST in Tasks 1 and 2 based on cough or breathing sounds with different spectrogram window lengths.}} 
\label{tab:lengthtuning}
\centering 
\resizebox{0.65\columnwidth}{!}{%
\begin{tabular}{lccc}
\toprule 
&  1024 & 2048 & 4096 \\
\midrule
Task 1 Cough & 0.86$\pm$0.03& \textbf{0.93}$\pm$\textbf{0.04} & 0.82$\pm$0.06   \\
Task 1 Breath & 0.85$\pm$0.05 & \textbf{0.94}$\pm$\textbf{0.04} & 0.84$\pm$0.06   \\
Task 2 Cough & 0.93$\pm$0.04 & \textbf{0.94}$\pm$\textbf{0.07} & 0.93$\pm$0.02   \\
Task 2 Breath & 0.89$\pm$0.07 & \textbf{0.93}$\pm$\textbf{0.04} & 0.92$\pm$0.07    \\
\bottomrule 
\end{tabular}}
\end{table}

\begin{table}[t]
\caption{Test F1 scores of HST in Tasks 1 and 2 based on cough or breathing sounds. Small, base, and large variants were considered.} 
\label{tab:sizetuning}
\centering 
\resizebox{0.65\columnwidth}{!}{%
\begin{tabular}{lcccc}
\toprule 
&  Small & Base & Large \\
\midrule
Task 1 Cough & 0.85$\pm$0.09& \textbf{0.93}$\pm$\textbf{0.04} & 0.86$\pm$0.03  \\
Task 1 Breath & 0.86$\pm$0.05 & \textbf{0.94}$\pm$\textbf{0.04} & 0.87$\pm$0.05  \\
Task 2 Cough & 0.89$\pm$0.05 & \textbf{0.94}$\pm$\textbf{0.07} & 0.89$\pm$0.05   \\
Task 2 Breath & 0.91$\pm$0.07 & \textbf{0.93}$\pm$\textbf{0.04} & 0.91$\pm$0.02   \\
\bottomrule 
\end{tabular}}
\end{table}

\revhl{The datasets examined here contain recordings from participants instructed to produce either cough or breathing sounds. Yet, there can be practical scenarios in which the modality of a given recording might be unknown. To assess the influence of modality labels on HST, we compared modality-specific and modality-agnostic models. Modality-specific models were trained and tested on a single known modality (either cough or breathing). Meanwhile, modality-agnostic models were trained and tested on a mixture of breathing and cough sounds without labels. Table \ref{tab:modalitytuning} lists the resultant performance metrics. The performance differences between modality-specific and modality-agnostic models are modest, suggesting that HST shows reasonable reliability against missing modality labels.}

\revhl{Finally, we examined the computational complexity of HST against a vanilla transformer variant that replaced local-windowed MSA layers with global MSA layers. Table \ref{tab:compload} lists total pre-training time on ImageNet, total training time and per-sample inference time on the Cambridge dataset, along with FLOPS, memory use, and number of parameters. Compared to the vanilla variant, HST requires lower FLOPS, memory use and fewer parameters to process spectrograms, and so it offers faster pre-training, training and inference.}

\begin{table}[t]
\caption{Test F1 scores of HST in Tasks 1 and 2 based on cough or breathing sounds, without and with pre-training.} 
\label{tab:traintuning}
\centering 
\resizebox{0.575\columnwidth}{!}{%
\begin{tabular}{lccc}
\toprule 
&  No Pre-training & Pre-training \\
\midrule
Task 1 Cough & 0.77$\pm$0.08& \textbf{0.93}$\pm$\textbf{0.04}   \\
Task 1 Breath & 0.70$\pm$0.10 & \textbf{0.94}$\pm$\textbf{0.04}   \\
Task 2 Cough & 0.83$\pm$0.07 & \textbf{0.94}$\pm$\textbf{0.07}    \\
Task 2 Breath & 0.83$\pm$0.07 & \textbf{0.93}$\pm$\textbf{0.04}    \\
\bottomrule 
\end{tabular}}
\end{table}

\begin{table}[t]
\caption{\revhl{Test F1 scores of HST in Tasks 1 and 2. Modality-specific models based on cough or breathing sounds, and modality-agnostic models based on a mixture of cough/breathing sounds are shown.}} 
\label{tab:modalitytuning}
\centering 
\resizebox{0.575\columnwidth}{!}{%
\begin{tabular}{lcccc}
\toprule 
 & Task 1 & Task 2 \\
\midrule
Modality: Cough & 0.93$\pm$0.04 & 0.94$\pm$0.07 \\
Modality: Breath & 0.94$\pm$0.04 & 0.93$\pm$0.04\\
Modality-agnostic & 0.91$\pm$0.04 &  0.92$\pm$0.04 \\
\bottomrule 
\end{tabular}}
\end{table}

\begin{table}[t]

\caption{\revhl{Complexity of HST and a vanilla transformer variant. Pre-training time, training time, per-sample inference time are listed along with FLOPS, memory use, and number of parameters.}} 
\label{tab:compload}
\centering 
\resizebox{0.925\columnwidth}{!}{%
\begin{tabular}{lcccccc}
\toprule
 & Pre-training & Training & Inference & FLOPS & Memory & N\textsubscript{params}\\
\midrule
HST & 114 h & 5.5 m & 19.4 ms & 8.8 G & 1.7 GB & 50 M\\
Vanilla  & 455 h & 13.7 m & 22.7 ms & 35 G & 3.6 GB & 171 M \\
\bottomrule
\end{tabular}}
\end{table}

\section{Discussion}
\revhl{Clinical testing of COVID-19 involves PCR and/or imaging procedures typically administered in centralized healthcare institutions} \cite{shi2021covid,9381578}. These cost-intensive procedures might not be broadly accessible in developing countries, and access might be delayed till relatively late stages of disease. In this context, remote monitoring via audio recordings of respiratory sounds can help pave the way to accessible preliminary screening \cite{ni2021automated}. Here, we introduce a hierarchical spectrogram transformer to detect COVID-19 with high accuracy from brief audio recordings of cough and breathing sounds. \revhl{Similar to manual auscultation, the proposed method does not serve as a diagnostic test. Yet, it holds promise as a preliminary assessment tool for potential COVID-19 cases that can assist in informed allocation of limited resources for timely testing and interventions \cite{Crotty2022}.} 

\revhl{Transformers can  capture contextual features in sound signals more effectively than CNN or RNN models \cite{gong2021ast}. Yet, vanilla transformers with global attention suffer from quadratic complexity \cite{dosovitskiy2020vit}, restricting their use under limited compute budgets on mobile devices. Instead, HST achieves high computational efficiency by leveraging local-windowed attention for linear complexity, and a hierarchical structure for progressively lowered spectrogram resolution. That said, analyses of respiratory sounds in this study were conducted offline for proof-of-concept demonstration. In practice, remote monitoring involves online processing of audio recordings  either on mobile devices or via communication with a cloud computing platform \cite{chowdhury2021qucoughscope, imran2020ai4covid}. To reduce computational and communication load, network distillation or hybrid CNN-transformer architectures might be adopted \cite{hinton2015distilling,transmsTMI}.}

\revhl{Here, demonstrations were performed on the Cambridge \cite{brown2020exploring} and COUGHVID \cite{orlandic2021coughvid} datasets. Across competing methods, we find generally higher detection performance on Cambridge versus COUGHVID. Several differences between the datasets might have contributed to this pattern. First, single audio recordings in Cambridge of duration 10.0$\pm$6.0 s contain repeated respiratory events (5 repeats for breathing, 3 repeats for cough), whereas those in COUGHVID of duration 6.6$\pm$2.5 s contain a single cough event per recording. Acoustic signatures of COVID-19 might manifest not only over short but also over longer time intervals, and prolonged recordings might facilitate separation of foreground-background signals. Thus, analyses based on multiple events per sample can improve detection sensitivity. Second, while Cambridge uses participant-reported COVID-19 status labels based on positive PCR test results, COUGHVID relies on expert-provided labels based on inspection of audio recordings. Thus, analyses on COUGHVID might suffer from higher label inaccuracy that can limit model performance. Third, native differences in microphone type, microphone positioning, recording environment, and compliance to recording instructions might have influenced the data quality, and thereby detection sensitivity. Future studies are warranted to identify ideal recording procedures for COVID detection from respiratory sounds.} 

\revhl{The Cambridge and COUGHVID datasets primarily comprise COVID and non-COVID groups (with or without cough symptoms), so the binary classification tasks examined here concerned segregation of these groups. While we find high detection accuracy for the COVID group against the non-COVID group with symptoms, a clinical characterization of respiratory disorders that underlie the cough symptoms is unavailable in the datasets. A practical concern regarding the adoption of a screening technology is its reliability on patient cohorts with varying types of disease \cite{liu2022deep,akay2022healthcare}. Recent studies suggest that respiratory sounds carry symptomatic cues to distinguish COVID-19 from other conditions such as asthma, chronic obstructive pulmonary disease (COPD), bronchitis and pertussis \cite{rudraraju2020cough,wheezes_crackles}. It remains important future work to examine whether HST can discriminate among a broad spectrum of respiratory diseases based on cough or breathing sounds alone. In cases where these two modalities do not carry sufficient information to detect COVID, speech or wheeze modalities might also be incorporated to boost model performance \cite{deshmukh2021interpreting,despotovic2021detection,hassan2020covid,perna2019deep}. Furthermore, additional clinical information such as patient demographics can be integrated to boost sensitivity of HST \cite{pal2021pay,pinkas2020sars}.} 

\revhl{Here, we find that HST outperforms competing methods based on analyses conducted on relatively modest-to-moderate sized datasets. An essential next step for validation of HST is demonstration of its reliability on broader patient cohorts. In practice, cross-checking the specific input features of an audio recording that drive the model output against known respiratory sound markers of COVID can be critical to avoid erroneous decisions when using black-box deep learning models. The explanatory analyses presented here indicate that the Grad-CAM algorithm applied on HST can identify important input features for COVID detection that closely match the expected discriminating features of respiratory sounds in COVID-19 \cite{wheezes_crackles}. Yet, it remains important future work to evaluate the efficacy of gradient-based explanatory methods such as Grad-CAM and uncertainty characterization methods in interrogating the decisions provided by HST.}

\section{Conclusion}
In this study, we proposed a deep learning approach to detect COVID-19 from respiratory sounds. A novel hierarchical transformer model, HST, was introduced to extract contextual features from spectrogram representations of audio signals. HST leverages local attention mechanisms over progressively growing windows to capture long-range context without excessive computational burden. While our demonstrations focused on COVID-19 screening, the contextual sensitivity of HST might also be helpful in detection of other pervasive respiratory disorders such as pneumonia, bronchitis or obstructive pulmonary disease.

\bibliographystyle{IEEEtran}
\bibliography{IEEEabrv,main}

\end{document}